\documentclass[letterpaper,11pt]{article}

\usepackage{fancyhdr}
\usepackage{amsmath}
\usepackage{amsbsy}
\usepackage{amssymb}
\usepackage{amsthm}
\usepackage{amscd}
\usepackage{amsfonts}
\usepackage{supertabular}
\usepackage{graphics}
\usepackage{graphicx}
\usepackage{verbatim}
\usepackage{subfigure}
\usepackage{epsfig}
\usepackage{xspace}
\usepackage{euscript}
\usepackage{alltt}
\usepackage{boxedminipage}
\usepackage{float}
\usepackage{times}
\usepackage[colorlinks]{hyperref}
\usepackage[square,numbers,compress]{natbib}
\usepackage{setspace}
\usepackage{epstopdf}
\usepackage{authblk}

\usepackage{rotating}
\usepackage{tikz}

\usepackage{algorithm}
\usepackage{algorithmic}

\def\Put(#1,#2)#3{\leavevmode\makebox(0,0){\put(#1,#2){#3}}}

\begin{document}

\title{Statistical characterization of microstructure evolution during compaction of granular systems composed of spheres with hardening plastic behavior}
\author[1,2]{Marcial~Gonzalez \thanks{Corresponding author. Tel.: +1 765 494 0904; fax: +1 765 496 7537.
E-mail address: marcial-gonzalez@purdue.edu (M. Gonzalez).}}
\author[1]{Payam~Poorsolhjouy \thanks{E-mail address: ppoorsol@purdue.edu (P. Poorsolhjouy).}}
\author[1]{Alex~Thomas}
\author[1]{Jili~Liu}
\author[1]{Kiran~Balakrishnan}
\affil[1]{School of Mechanical Engineering, Purdue University, West Lafayette, IN 47907, USA}
\affil[2]{Ray W. Herrick Laboratories, Purdue University, West Lafayette, IN 47907, USA}

\maketitle
\begin{abstract}

An extensive numerical campaign of particle mechanics calculations that predict microstructure formation and evolution during die compaction, up to relative densities close to one, of monodisperse plastic spheres that exhibit power-law plastic hardening behavior is presented. The study is focused on elucidating the relationship between particle plastic properties, loading conditions, and statistical features of the resulting microstructure. This communication provides fundamental insight into the achievable space of microstructures through die compaction, for given plastic stiffness and hardening exponent at the particle scale.
\end{abstract}


\section{Introduction}

Granular materials have distinct characteristic features that arise from their discrete nature. As opposed to homogeneous continuous materials, these materials are heterogeneous and disordered. Their load bearing properties, therefore, depend heavily on their microstructure \cite{ChangMisra-ElastoPlastic,Payam-2d}. Compaction of granular materials is of great interest to soil mechanics as well as powder compaction. It is then the microstructure of the highly confined soil or dense powder which dictates the properties and performance of the system. Therefore, the elucidation of the relationship between particle properties, granular morphology, loading conditions, and microstructural features developed during compaction is desirable for fundamentally understanding the achievable design space map, which in turn allows for optimizing performance. Mechanistic continuum models, such as such as the granular micromechanics approach \cite{Poorsolhjouy-2018, Payam-BandGap, Payam-Identification, Payam-LoadPath,Payam-Gibbs}, and discrete models, such as the particle mechanics approach \cite{Gonzalez-2016,yohannes2016evolution,yohannes2017discrete,yohannes2017discretenumerical,gonzalez2017generalized}, capable of describing strength formation and microstructure evolution during the compaction process are of great interest. In this communication, we restrict attention to die compaction of monodisperse plastic spheres that exhibit power-law plastic hardening behavior. Specifically, we investigate the achievable design space map between particle plastic properties, loading conditions, and statistical features of the resulting microstructure using an extensive set of particle mechanics simulations. Next, we briefly describe the particle mechanics approach to powder compaction \cite{Gonzalez-2016}.

The particle mechanics approach for granular systems under high confinement, developed by Gonzalez and Cuiti\~{n}o \cite{Gonzalez-2016}, describes each individual particle in the powder bed, and the collective rearrangement and deformation of the particles that result in a quasi-statically compacted specimen. An equilibrium configuration is therefore defined by the solution of a system of nonlinear equations that corresponds to static equilibrium of the granular system, that is sum of all contact forces acting on each particle equals zero. Contact forces are defined as functions of relative displacement between neighbor particles $i$ and $j$, $\gamma_{ij}=R_i+R_j-||\textbf{x}_i-\textbf{x}_j||$ where $R_i$ and $R_j$ are the radii of the two particles and $\textbf{x}_i$ and $\textbf{x}_j$ denote their locations. This approach has been used to predict the microstructure evolution during die-compaction of elastic spherical particles up to relative densities close to one. By employing a nonlocal contact formulation that remains predictive at high levels of confinement \cite{Gonzalez-2012}, this study demonstrated that the coordination number depends on the level of compressibility of the particles and thus its scaling behavior is not independent of material properties as previously thought. The study also revealed that distributions of contact forces between particles and between particles and walls, although similar at jamming onset, are very different at full compaction--noting that particle-wall forces are in remarkable agreement with experimental measurements reported in the literature.

In this communication, we extend the study to die-compaction of monodisperse plastic spheres that exhibit power-law plastic hardening behavior. For context, we first recall that Hertz contact theory for elastic spherical particles is based on Hooke's law of linear elasticity \cite{timoshenko1970theory}. Specifically, the contact force $F$ between two identical spherical particles with radius $R$, elastic modulus of $E$ and Poisson's ratio of $\nu$, is given by  
\begin{equation}
	F
	=
	n_\mathrm{H}(\gamma)_+^{3/2} 
	\ \ \ \text{where } \ \ \ 
	n_\mathrm{H}
	=
	E (2 R)^{1/2}/3(1-\nu^2)\\
\label{Hertz-elastic}
\end{equation}
where $( \cdot )_+ = \max\{\cdot, 0\}$. For plastic spherical particles, we adopt a rigid plastic flow formulation and assume a power-law plastic hardening behavior, i.e., $\sigma=\kappa\epsilon^{1/m}$ where $\kappa$ is the plastic stiffnesses and $m \ge 1$ is the plastic law exponent or hardening exponent---the behavior reduces to linear once $m=1$. Specifically, the contact force $F$ for two identical plastic spheres  \cite{Storakers-Plastic} with radius $R$, plastic stiffness $\kappa$ and hardening exponent \textit{m}, is given by
\begin{equation}
	F
	=
	n_\mathrm{P} \left(c^2 R \gamma\right)_+^{1+1/2m}
	\ \ \ \text{where } \ \ \ 
	n_\mathrm{P}
	=
	\pi k R^{-1/m} \kappa
\label{f-plastic}
\end{equation}
with  $k=3\times 6^{-1/m}$ and $c^2 = 1.43~\mathrm{e}^{-0.97/m}$.
We note here that the plastic stiffness $\kappa$ is a scaling factor and does not affect the compacted granular system 's behavior qualitatively. In sharp contrast, the hardening exponent $m$ affects material behavior and microstructure evolution \cite{yohannes2016evolution,yohannes2017discrete,yohannes2017discretenumerical} and will thus be the central focus of this work. It is worth noting that the systematic development of nonlocal contact formulations for elasto-plastic spheres is a worthwhile direction of future research and, though beyond the scope of this work, is currently being pursued by the author (see \cite{Gonzalez-2012, Gonzalez-2016} for an elastic nonlocal contact formulation).

\begin{figure}[htb]
	\centering
	\includegraphics[scale=1.25]{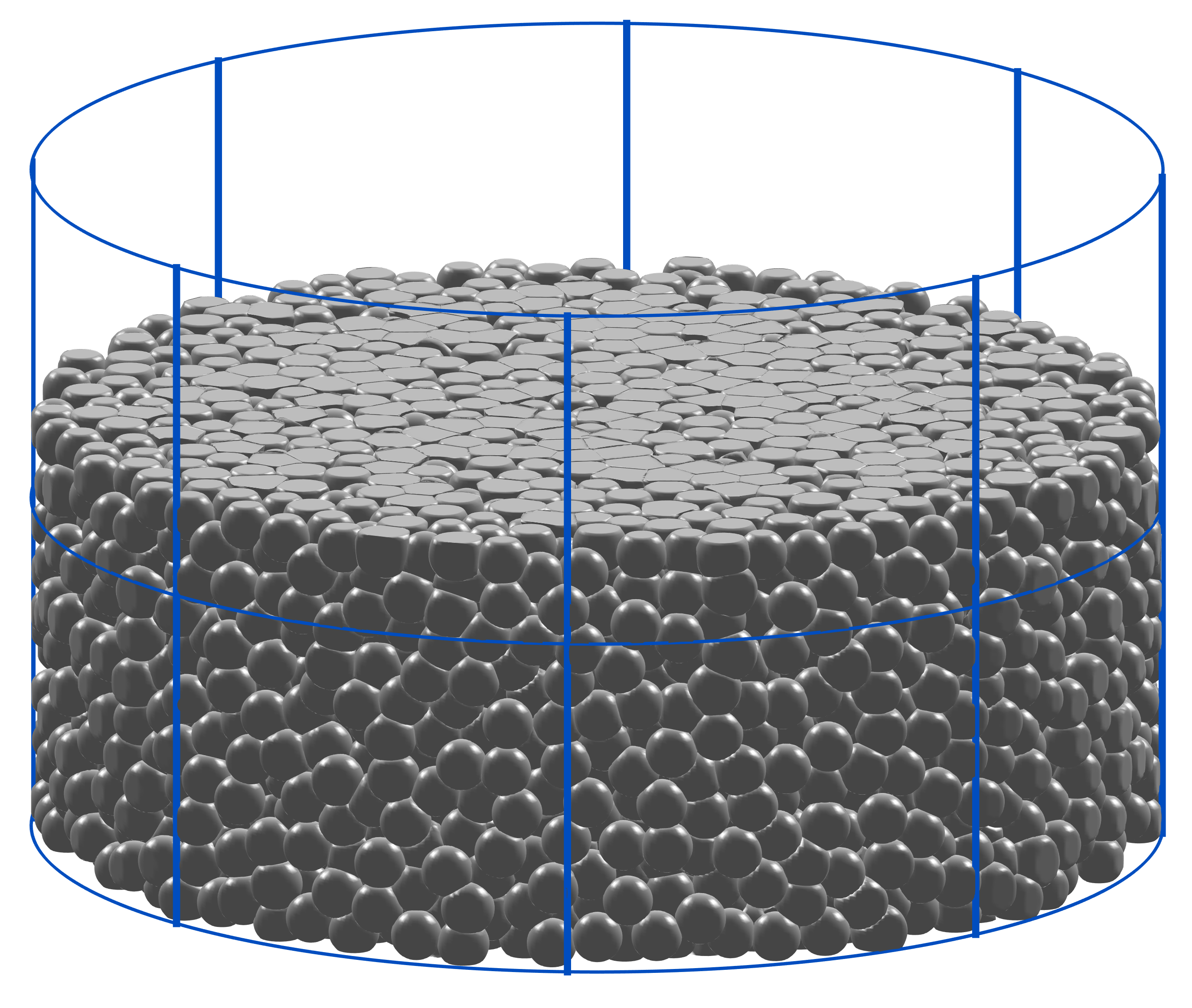}
	\caption{Three-dimensional representation of the 6k granular packing compacted at relative density $\rho= 0.95$.}
	\label{Fig-3DPacking}
\end{figure}

We specifically study noncohesive frictionless granular systems with three different number of particles, namely $6,512$, $22,180$ and $39,914$---which are refereed to as $6$k, $20$k and $40$k packings, respectively. The weightless spherical particles have $\kappa=245$~GPa and $m\in [1,17]$. The granular bed, which is numerically generated by means of a ballistic deposition technique, is constrained by a rigid cylindrical die of diameter $D$. Diameter and particle radius are different for each packing, namely 6k ($D=10$~mm, $R=220~\mu$m), 20k ($D=15$~mm, $R=220~\mu$m), and 40k ($D=10$~mm, $R=125~\mu$m). Therefore, this study will address the effect on material behavior and microstructure evolution of, not only changing m, but also particle dimensions and $D/R$ ratio. Figure~\ref{Fig-3DPacking} shows a three-dimensional representation of a compacted granular packing.

\section{Results and Discussions}

We next study the microstructure formation and evolution of monodisperse plastic spheres during die compaction. We investigate the evolution of statistical features of the jamming point, the mechanical coordination number, the applied compaction pressure, and the distribution of magnitude and orientation of contact forces and radii. Since these results are sensitive to the numerical tolerance and boundary effects, we adopt the findings reported in \cite{Gonzalez-2016} for the interparticle deformation above which a contact force is identified as non-zero, i.e., $\epsilon_\mathrm{tol}=\gamma/2R=0.005\%$, and for the distance between the boundary and bulk region, i.e., Gap~$=4R$. In addition, error bounds in the plots represent the 95$\%$ confidence envelope for fitted coefficients.

\subsection{Jamming point and mechanical coordination number}

\begin{figure*}[htb]
\centering{
\begin{tabular}{lll}	
	\includegraphics[trim= 8mm 0mm 85mm 0mm, clip, scale=0.62]{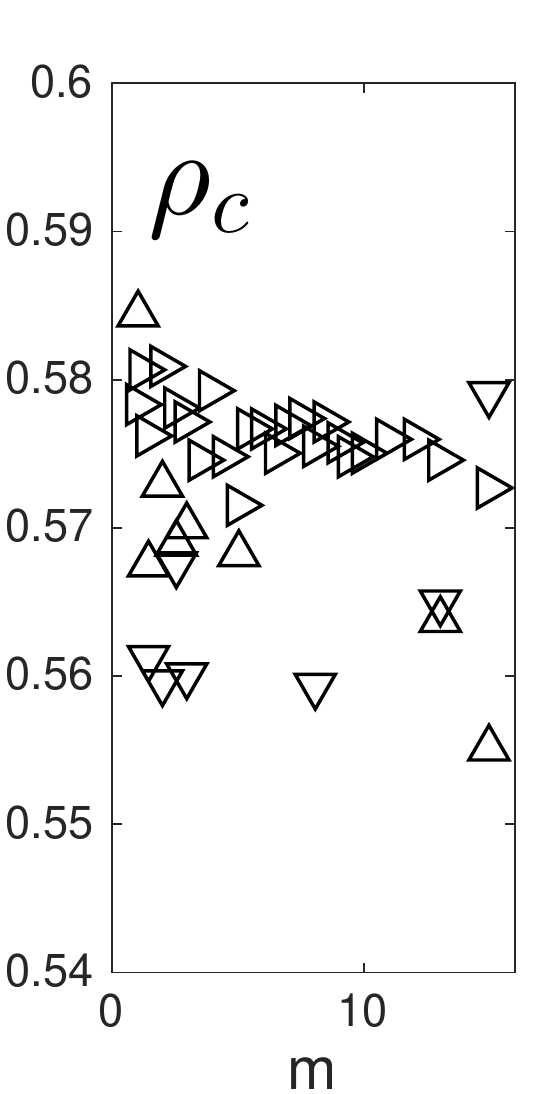}
	&
	\includegraphics[trim= 8mm 0mm 85mm 0mm, clip, scale=0.62]{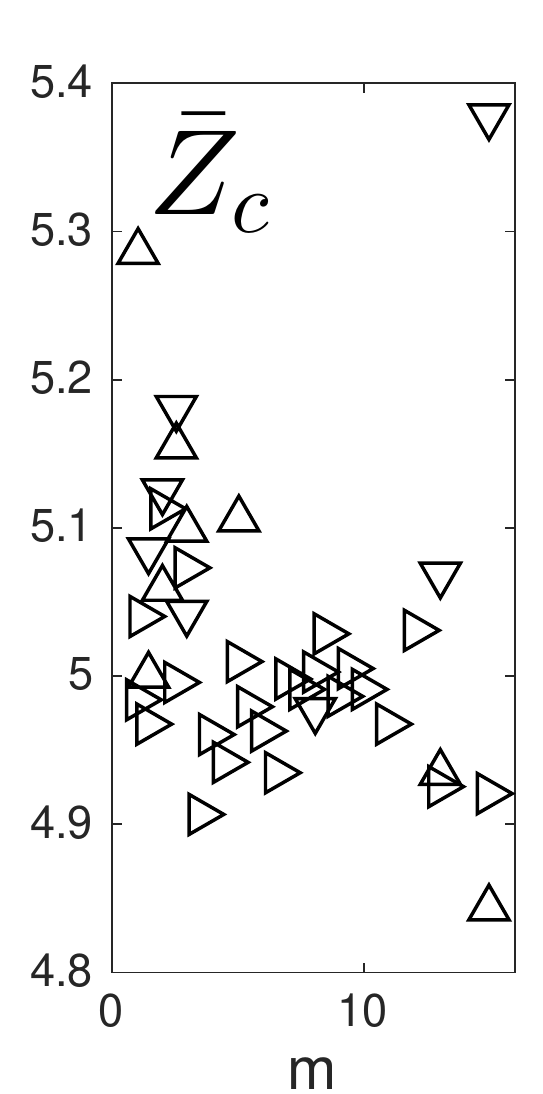}
	&
	\includegraphics[trim= 8mm 0mm 85mm 0mm, clip, scale=0.62]{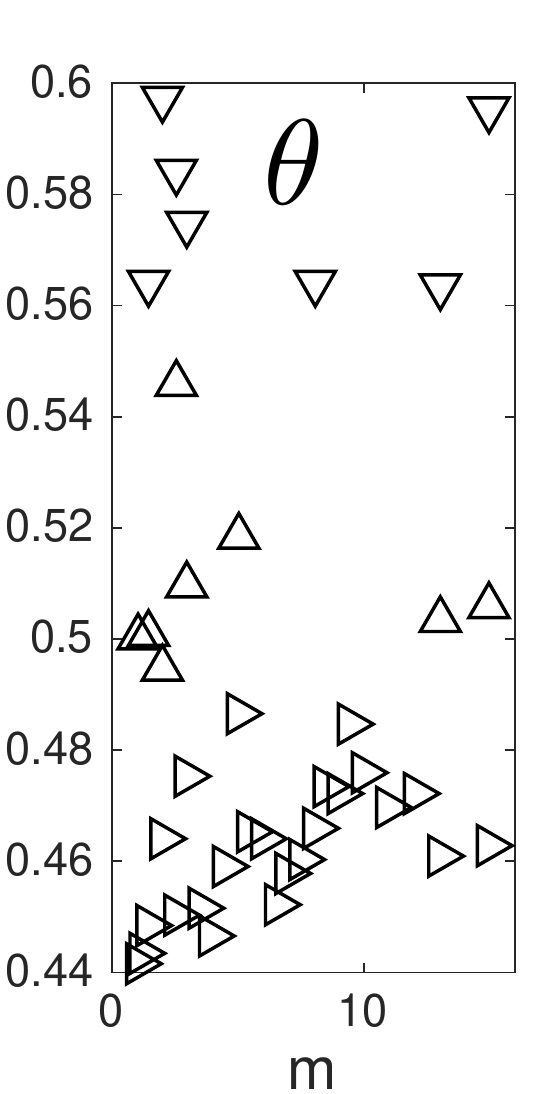}
	\vspace{-.17in}
\\
	\small{(a)}
	&
	\small{(b)}
	&
	\small{(c)}
\end{tabular}
\vspace{-.1in}
\begin{tabular}{ll}	
	\includegraphics[trim= 8mm 0mm 54mm 0mm, clip, scale=0.62]{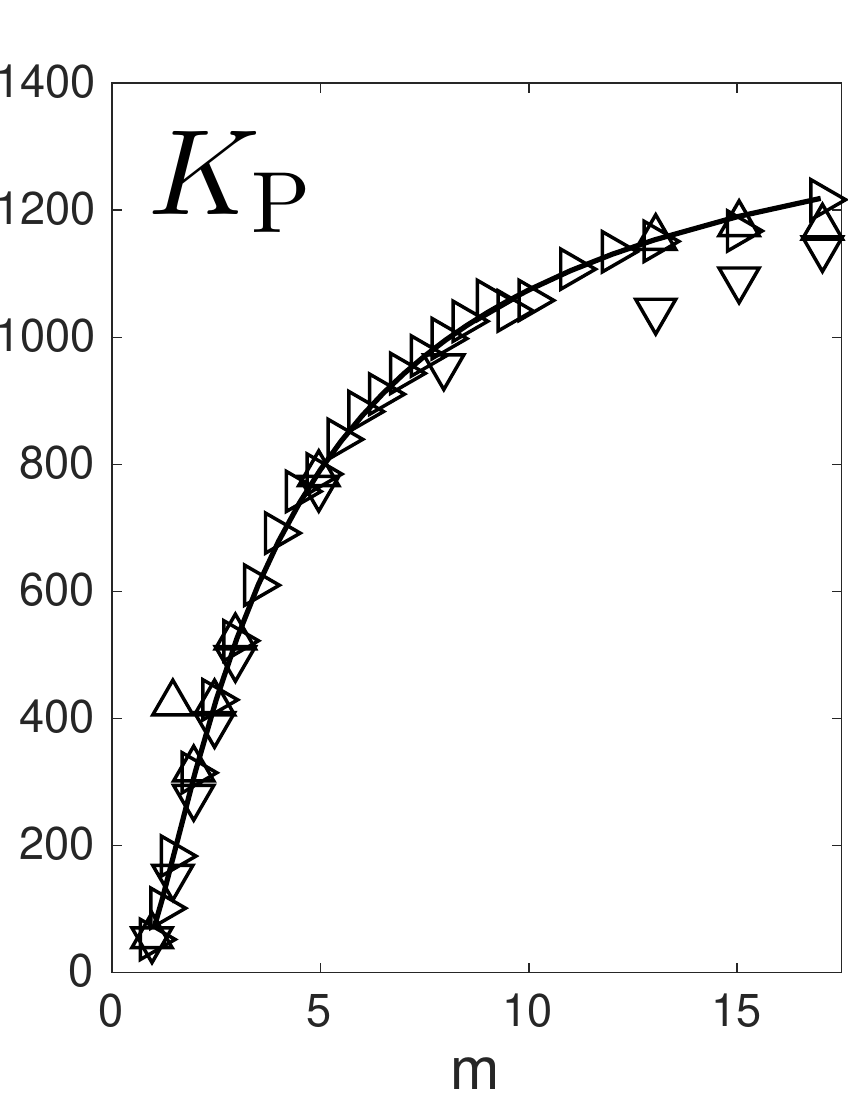}
	&
	\includegraphics[trim= 8mm 0mm 54mm 0mm, clip, scale=0.62]{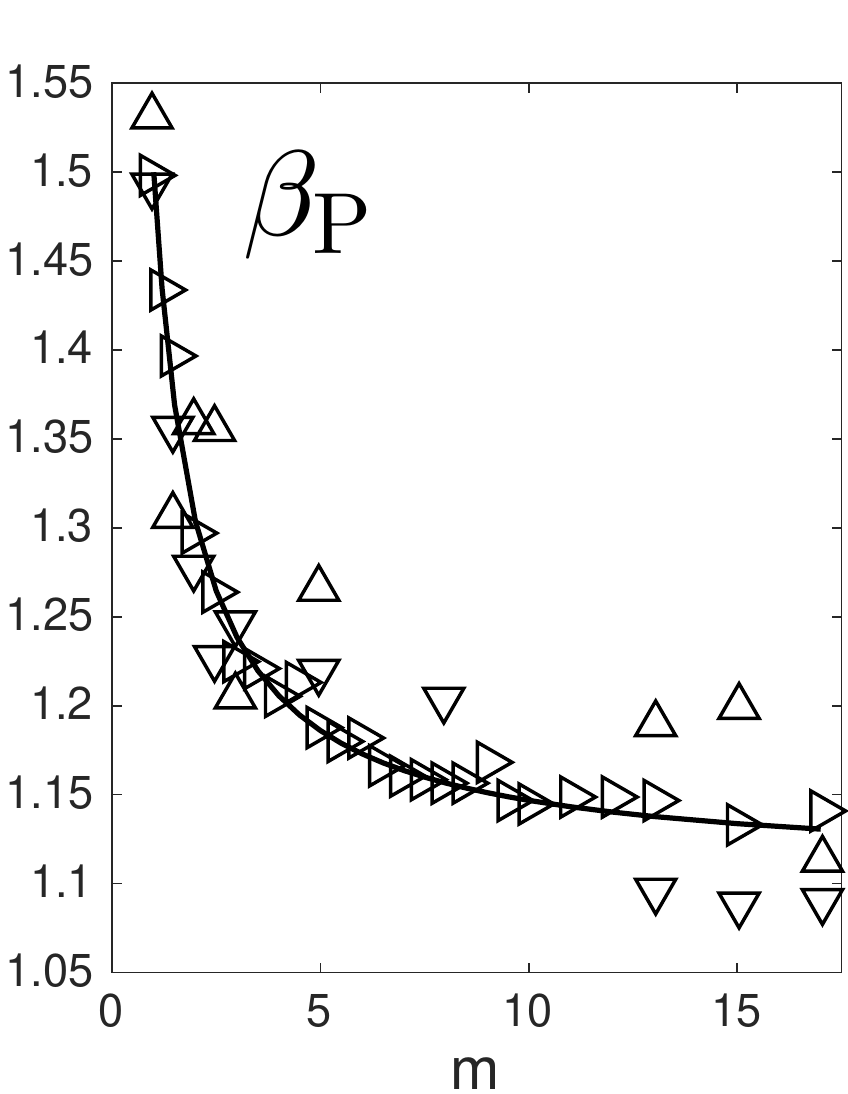}
	\vspace{-.17in}
\\
	\small{(d)}
	&
	\small{(e)}
\end{tabular}
}
\caption{Critical relative density $\rho_c$, minimal mean coordination number $\bar{Z}_c$, critical exponent $\theta$, and compaction pressure parameters $\beta_\mathrm{P}$ and $K_\mathrm{P}$ [MPa], for different hardening exponents $m$ and packings (6k $\rhd$, 20k $\vartriangle$, 40k $\triangledown$). The symbols in (a)-(c) correspond to the best fit of equation \eqref{EQ-Z vs rho} to the mean coordination number obtained from the particle contact mechanics simulation of the granular bed. The symbols in (d)-(e) correspond to the best fit of equation \eqref{Eq.s-rho-Plastic} to the punch compaction pressure obtained from the simulation.}
\label{Fig-Z-Pressure-vs-m}
\end{figure*}

Compaction is initially characterize by particle rearrangement and very small deformation, and it is followed by a jamming transition occurring at a critical relative density $\rho_c$ \cite{Oda-RelativeDensity,Salot-RelativeDensity}. After jamming, macroscopic deformation is achieved by increasingly larger particle deformations and mean mechanical coordination number $\bar{Z}$ \cite{MakseForceChain,Radjai}---mean number of non-zero contact forces between a particle and its neighbors. The evolution of $\bar{Z}$ during compaction is given by
\begin{equation}
	\bar{Z} - \bar{Z_c} 
	= 
	\bar{Z_0} (\rho -\rho_c)^\theta
\label{EQ-Z vs rho}
\end{equation}
where $\bar{Z}_c$ is the minimal mean coordination number, $\theta$ is the critical exponent, and $\bar{Z}_0$ is constant of proportionality. These parameters that characterize the jamming transition and mean coordination number evolution can be determined from the particle mechanics simulations (see \cite{Gonzalez-2016} for details about the procedure). Figure~\ref{Fig-Z-Pressure-vs-m}(a)-(c) shows the values obtained from fitting Eq.~\eqref{EQ-Z vs rho} to 45 simulations. It is evident from figure that $\rho_c$ does not vary significantly by changing material properties and it is rather a structural or geometric property of the packing---there is naturally a weak dependency on $D/R$ but, for the studied packings, $\rho_c$ is within 0.55 to 0.59. Furthermore, even though there is a weak dependency of $\bar{Z}_c$ on hardening exponent, the critical exponent $\theta$ does not depend on material properties, confirming that the evolution of the mean coordination number is mainly a geometric behavior of the granular system.

\subsection{Applied compaction pressure}

The compaction pressure applied by the upper and lower punches $\sigma$ and the reaction at the die wall are effective macroscopic responses to the deformation process that are predicted from the particle mechanics simulations. For packings of monodisperse elastic spheres under die compaction and Hertzian interactions \cite{Gonzalez-2016}, $\sigma$ follows a power-law function of relative density $\rho$
\begin{equation}
\sigma = K_\mathrm{H}(\rho - \rho_c)^{\beta_\mathrm{H}}
\label{Eq.s-rho-Elastic}
\end{equation}
where $K_\mathrm{H} \approx E/2\pi(1-\nu^2)$ and $ \beta_H \approx 3/2$ does not depend on material properties. For packings of monodisperse plastic particles, the stress follows a similar power-law function of relative density, that is 
\begin{equation}
\sigma = K_\mathrm{P}(\rho-\rho_c)^{\beta_\mathrm{P}}
\label{Eq.s-rho-Plastic}
\end{equation}
However, in sharp contrast to the Hertzian case, the exponent $\beta_\mathrm{P}$ depends on material properties---as does the plastic contact law's exponent in equation Eq. \ref{f-plastic}. Figure~\ref{Fig-Z-Pressure-vs-m}(d)-(e) shows the values obtained from fitting Eq.~\eqref{Eq.s-rho-Plastic} to 45 simulations. It is evident from the figure that both $K_\mathrm{P}$ and $\beta_\mathrm{P}$ depend on the hardening coefficient $m$, but only weakly on $R$ and $D/R$. We propose the following expressions for these functions (Fitted curves in Figure \ref{Fig-Z-Pressure-vs-m})
\begin{equation} 
	\beta_\mathrm{P}
	=
	\beta_\mathrm{P(\infty)}+(\beta_\mathrm{P(1)}-\beta_\mathrm{P(\infty)})/m
\label{EQ-BetaP}
\end{equation}
\begin{equation}
	K_\mathrm{P} 
	= 
	K_\mathrm{P(1)}^{1/m} K_\mathrm{P(\infty)}^{1-1/m}
\label{EQ-KP}
\end{equation}	
where $\beta_\mathrm{P(1)}=3/2$, $\beta_\mathrm{P(\infty)}=1.11$, $K_\mathrm{P(1)}=64.9$~MPa and $K_\mathrm{P(\infty)}=1462$~MPa are $\beta_\mathrm{P}$ and $K_\mathrm{P}$, at $m=1$ and $m \rightarrow \infty$.

\subsection{Probability distribution of contact forces}

The probability distribution of normalized contact forces $\bar{f}=f/f_{av}$ in monodisperse granular systems of plastic spheres with different hardening exponent $m$ is proposed to follow
\begin{equation}
	P(\bar{f};m) 
	\propto 
	\bar{f}^{\frac{3(\gamma -1)}{2+1/m}} 
	\exp
	\left[ 
	-\lambda^\alpha \left|\bar{f}^{\frac{3}{2+1/m}}-f_0\right|^\alpha
	\right]
\label{Eq-Force}
\end{equation}
where $\gamma$ is a shape parameter that controls the behavior of weak forces, $\alpha$ is the exponent of algebraic tail that controls large forces, and $f_0$ is a translation factor. It should be noted that for $m=1$, the function simplifies to the elastic behavior \cite{Gonzalez-2016}. The probability distribution function is defined such that its integral over all force values, $\int_{0}^{+\infty}P(\bar{f})\mathrm{d}\bar{f}$, remains finite provided that $\alpha>1$ and $\gamma>0$. It should be noted that the probability distribution function of particle-particle contacts and particle-wall contacts are studied separately and normalized  by their corresponding average force values, $f_{av}$. 

\begin{figure}[htb]
\centering{
	\begin{tabular}{c}
	\includegraphics[trim= 13mm 2mm 0mm 0mm, clip, scale=0.52]{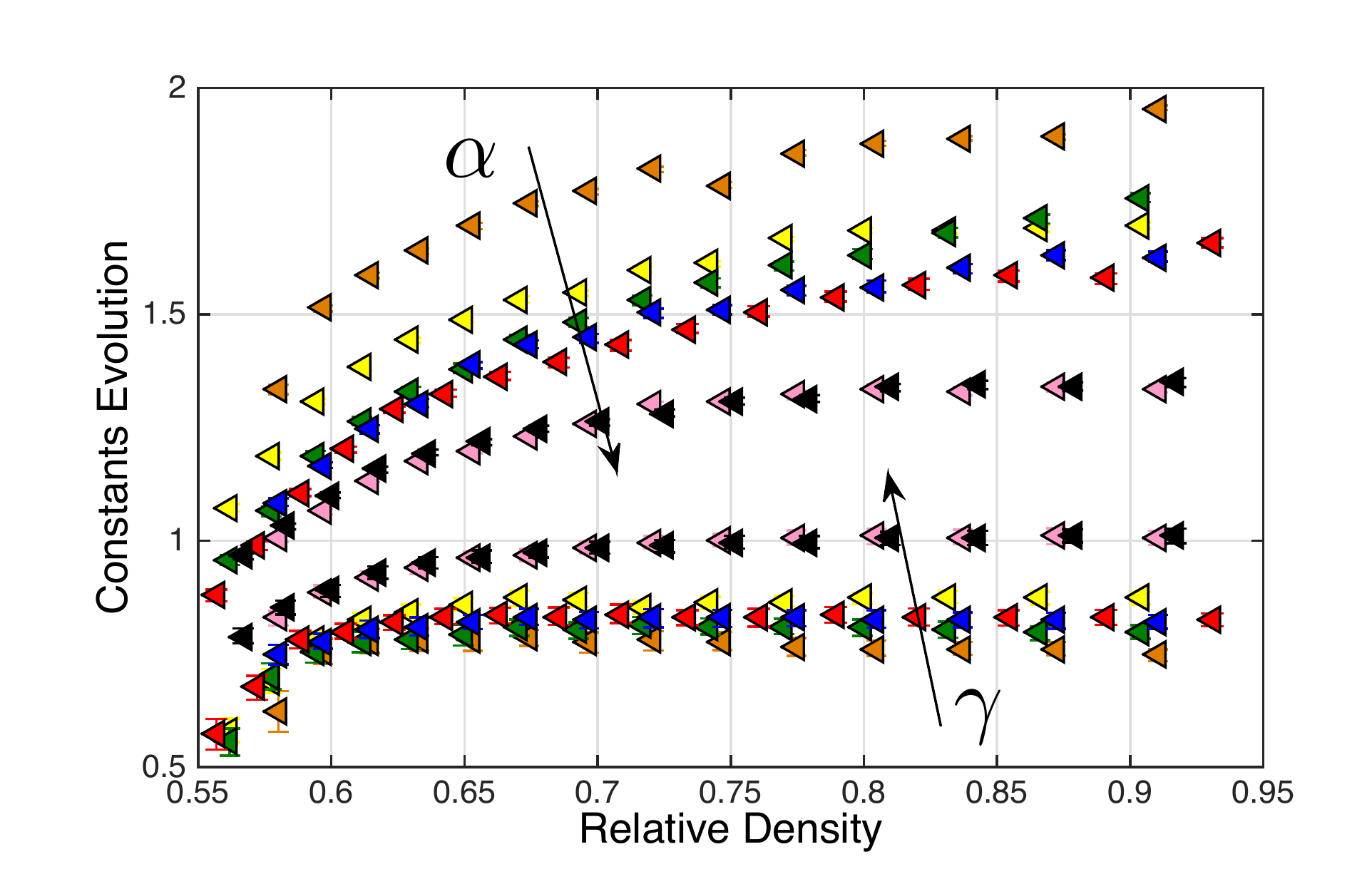} 
	\end{tabular} 
}
\caption{Statistical description of the probability distribution of particle-particle contact forces in 40k packings of plastic spheres with hardening exponent $m$ equal to 1, 1.5, 3, 5, 8, 13, and 15---the arrows indicate directions of increasing $m$.}
\label{Fig-p-p-parameters}
\end{figure}

\begin{figure}[htb]
\centering{
	\includegraphics[trim= 0mm 0mm 0mm 0mm, clip, scale=0.52]{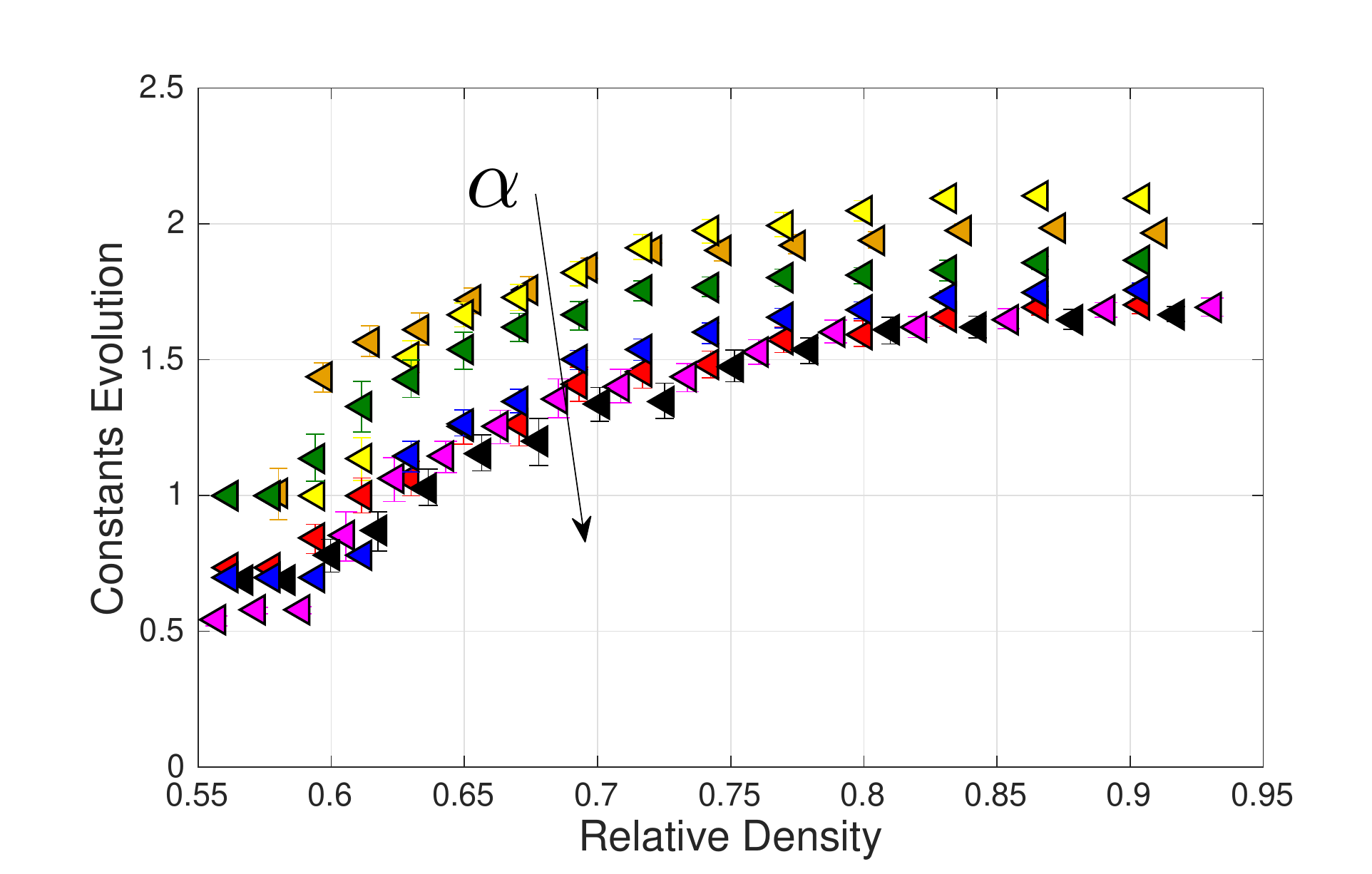} 
	\\
	\includegraphics[trim= 0mm 0mm 0mm 0mm, clip, scale=0.52]{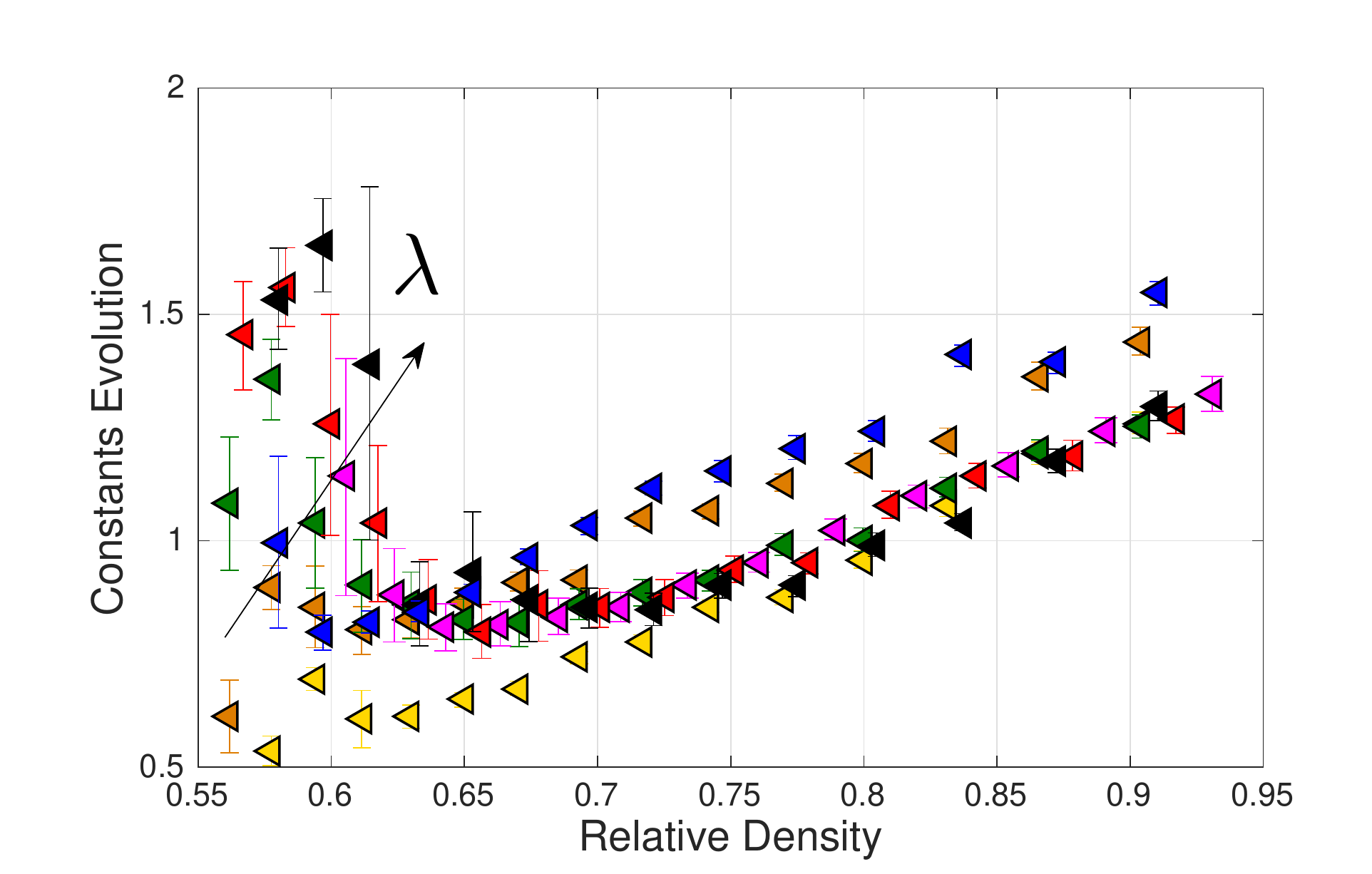}
}
\caption{Statistical description of the probability distribution of particle-wall contact forces in 40k packings of plastic spheres with hardening exponent $m$ equal to 1, 1.5, 3, 5, 8, 13, and 15---the arrow indicates direction of increasing $m$.}
\label{Fig-p-w-Alpha.Lambda}
\end{figure}

The proposed probability distribution of contact forces has four parameters, namely $f_0$, $\alpha$, $\lambda$, and $\gamma$. These parameters, however, are interrelated. For particle-particle contacts, the independent parameters are $\alpha$ and $\gamma$ which evolve as functions of relative density during compaction as shown in Figure \ref{Fig-p-p-parameters}. These values are the best fit of the particle mechanics results to Eq. \eqref{Eq-Force}, adopting
\begin{equation} 
	f_0
	=
	\sqrt{\alpha-(2+1/m)/3}
	\hspace{1mm};\hspace{1cm} 
	\lambda
	=
	\sqrt{\alpha}-1/2
\label{Eq-p-p-param}
\end{equation}
Similarly, for particle-wall contacts, the independent parameters are $\alpha$ and $\lambda$ which evolve as functions of relative density during compaction as shown in Figure \ref{Fig-p-w-Alpha.Lambda}. These values are also the best fit of the particle mechanics results to Eq. \eqref{Eq-Force}, now adopting
\begin{equation}
	f_0
	=
	\alpha-(2+1/m)/3
	\hspace{1mm};\hspace{1cm} 
	\gamma
	=
	2-(2+1/m)/3
\label{Eq-p-w-param}
\end{equation}

\begin{sidewaysfigure}
\centering{
\begin{tabular}{llll}
	\includegraphics[scale=0.61]{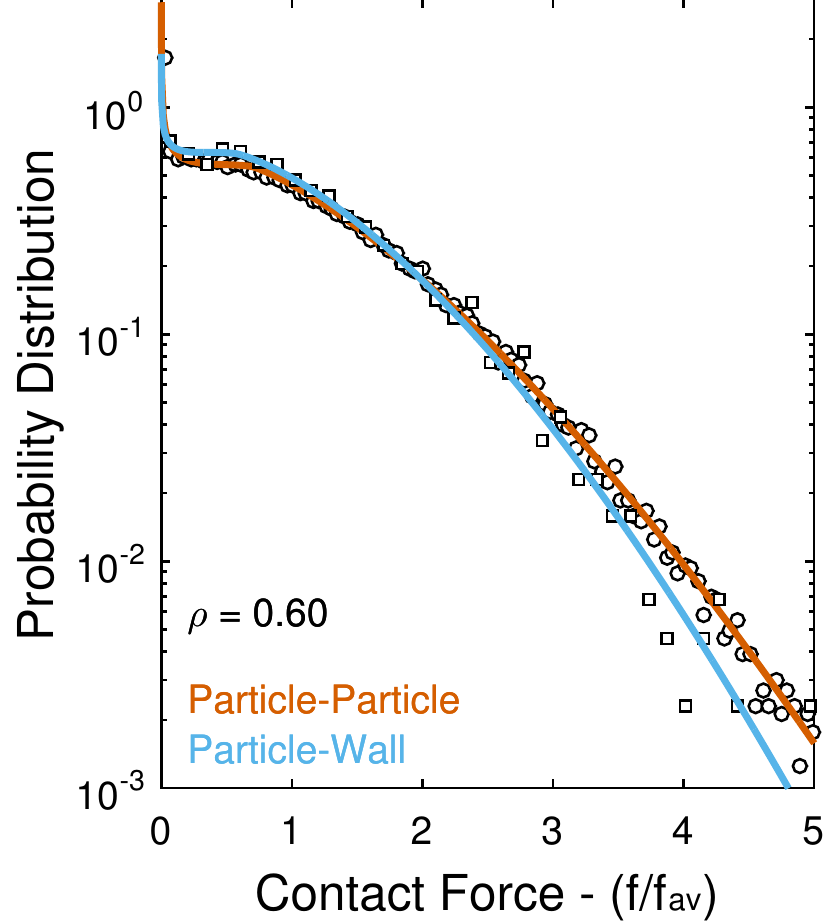}
	&
	\includegraphics[scale=0.61]{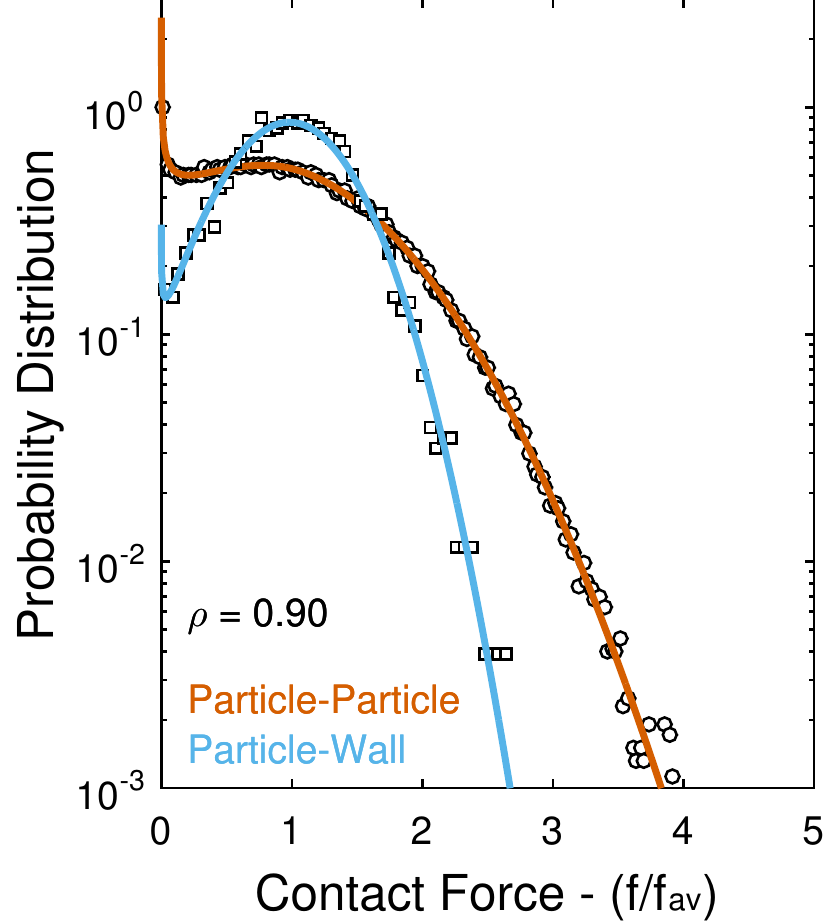}	
	&
	\includegraphics[scale=0.61]{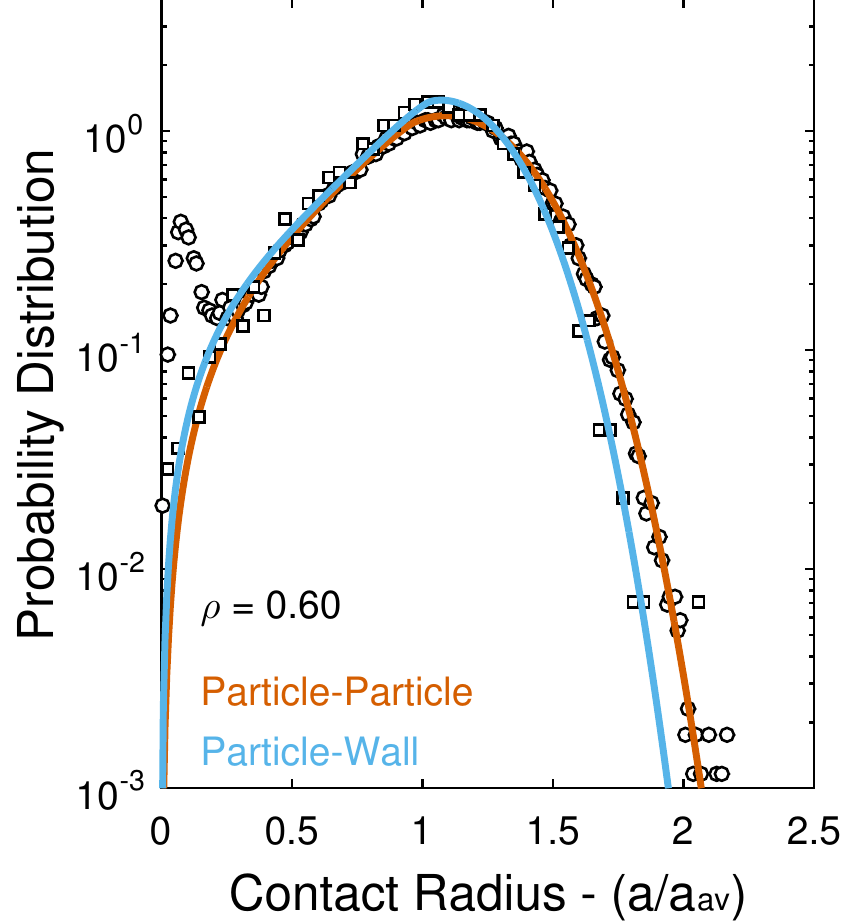}
	&
	\includegraphics[scale=0.61]{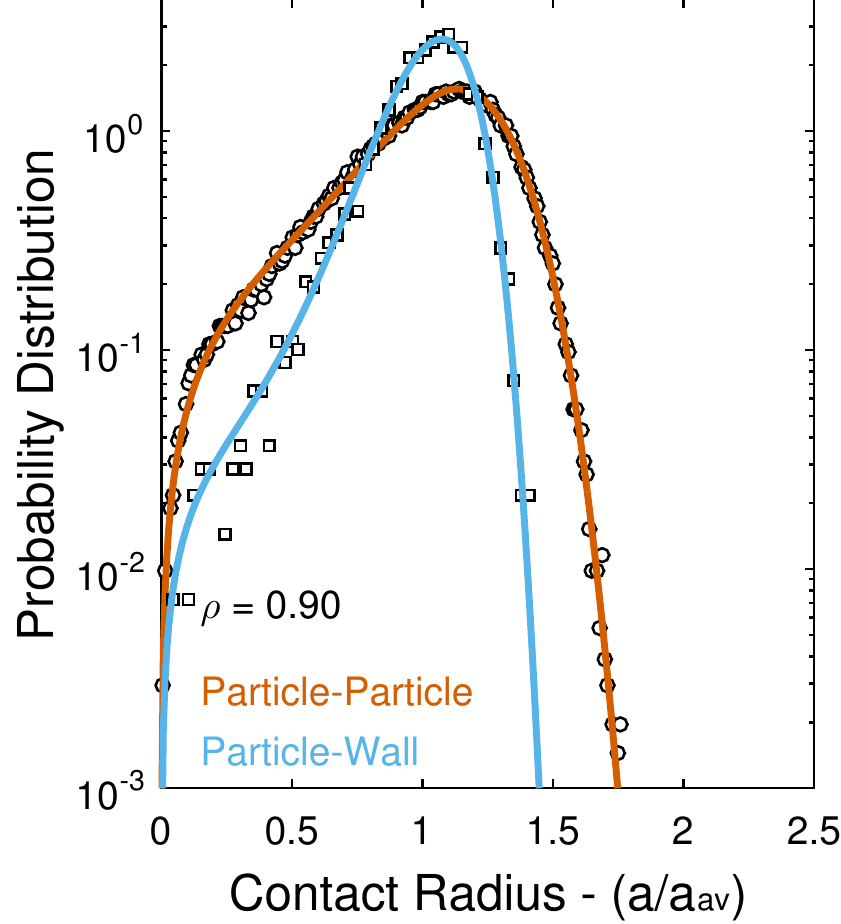}	
	\vspace{-.17in}
\\
	\small{(a)}
	&
	\small{(b)}
	&
	\small{(c)}
	&
	\small{(d)}
\\
	\includegraphics[scale=0.61]{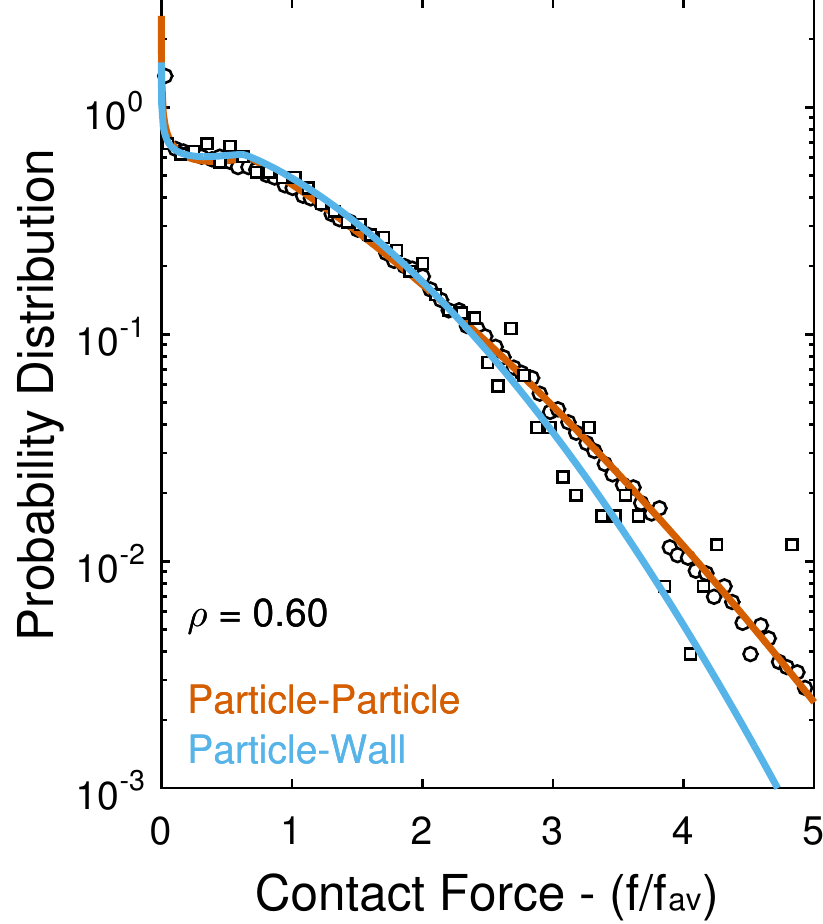}
	&
	\includegraphics[scale=0.61]{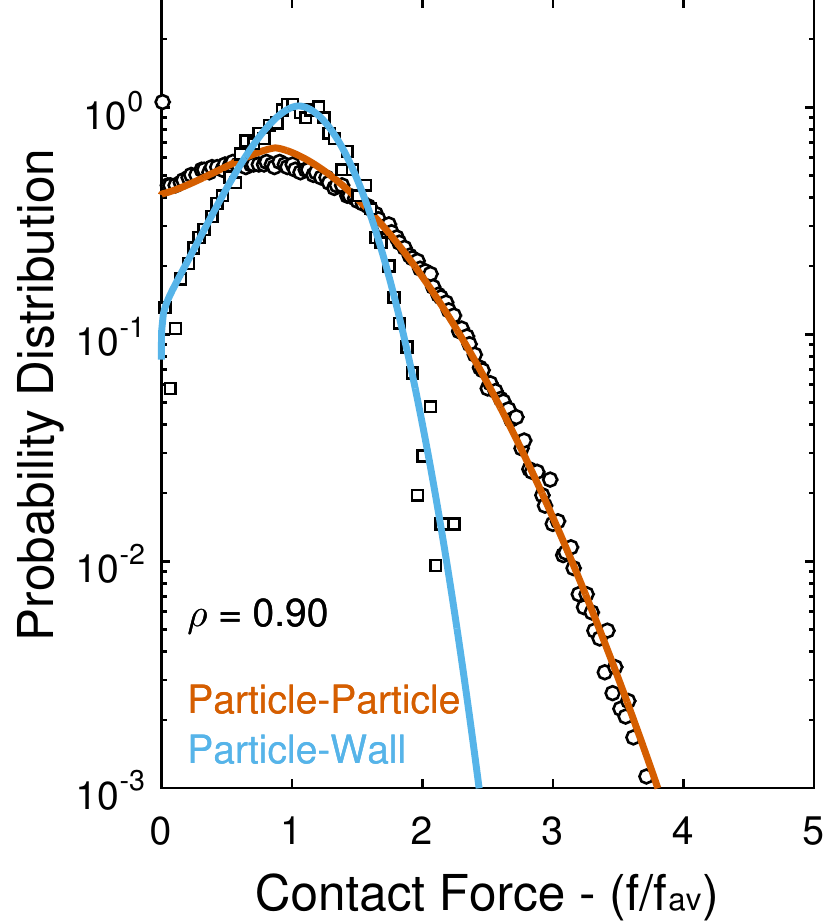}	
	&
	\includegraphics[scale=0.61]{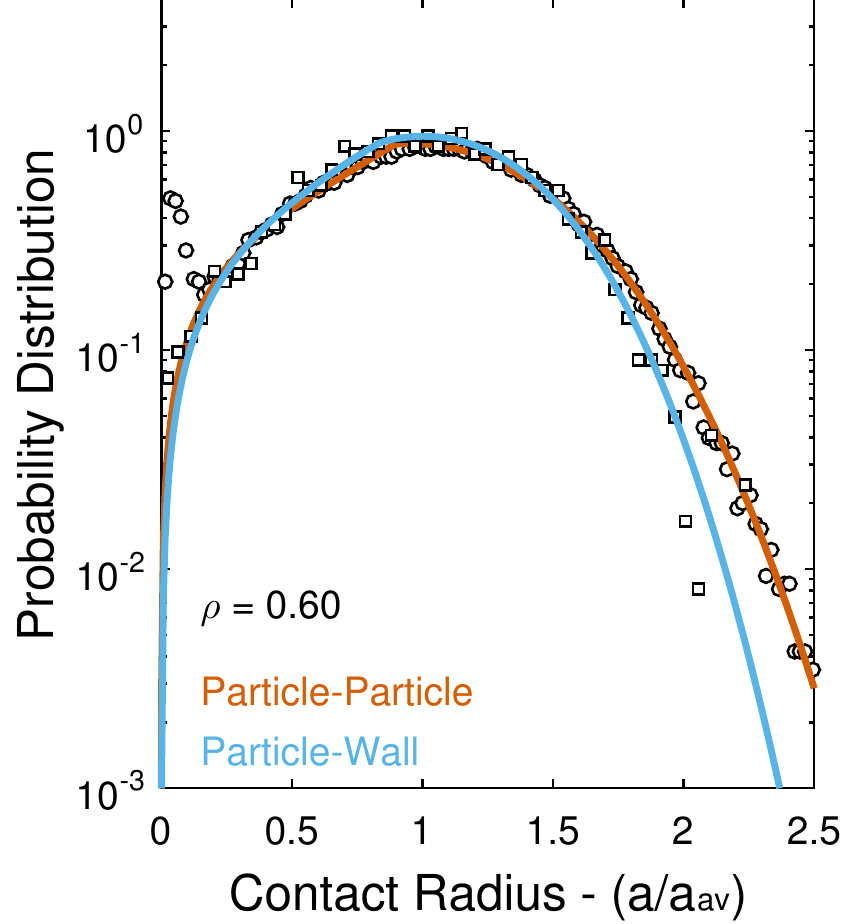}
	&
	\includegraphics[scale=0.61]{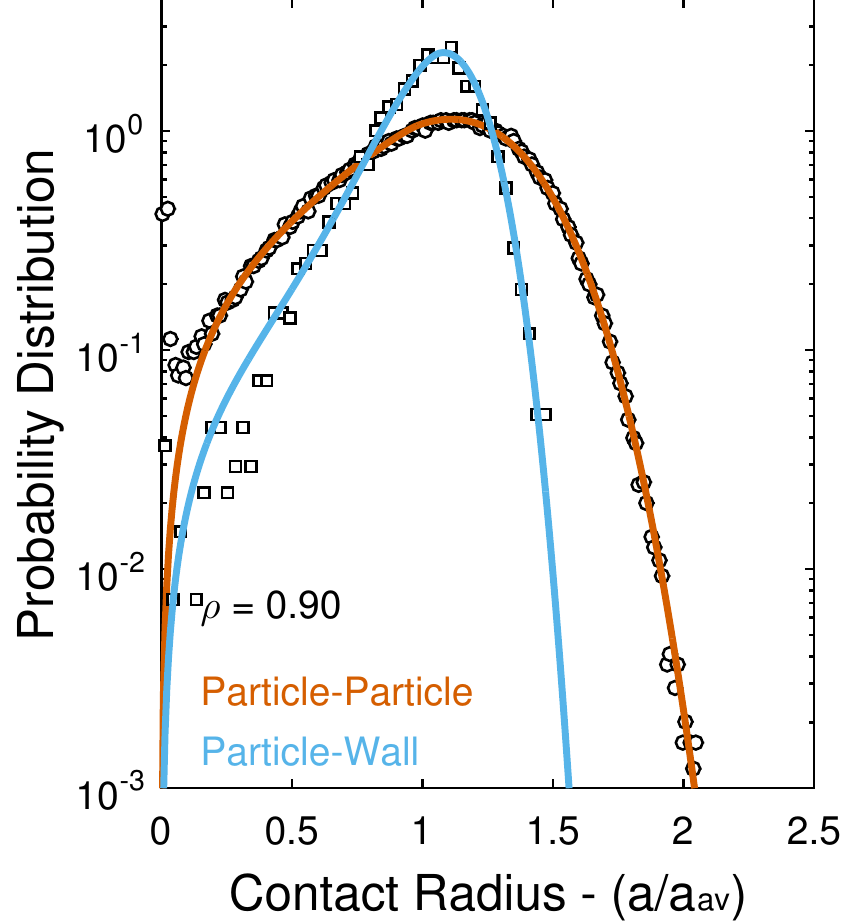}		
	\vspace{-.17in}
\\
	\small{(e)}
	&
	\small{(f)}
	&
	\small{(g)}
	&
	\small{(h)}
\end{tabular}
}
\caption{Probability distribution of particle-particle ($\circ$) and particle-wall ({\scriptsize $\square$}) contact forces (a,d,e,f) and radii (c,d,g,h) determined from particle mechanics simulations of 40k packings of plastic spheres with hardening exponent $m=1$ (top row) and $m=5$ (bottom row). Solid lines correspond to the best fit of numerical results to \eqref{Eq-Force}, with \eqref{Eq-p-p-param}-\eqref{Eq-p-w-param}, and to \eqref{Eq-Radius} for contact forces and radii, respectively.}
\label{Fig-ForcesRadii-m=1AND5}
\end{sidewaysfigure}

Figure \ref{Fig-ForcesRadii-m=1AND5} shows that Eq. \eqref{Eq-Force} and relationships \eqref{Eq-p-p-param}-\eqref{Eq-p-w-param} accurately describe the probability distributions obtained from particle mechanics simulations. It is also evident from Figures \ref{Fig-p-p-parameters} and \ref{Fig-ForcesRadii-m=1AND5} that, although similar at $\rho_c$, the evolution during compaction of these distributions depends on the hardening exponent $m$. Specifically, the tails of both distributions transition from exponential ($\alpha=1$) to algebraic, being Gaussian ($\alpha=2$) only for $m=1$ (cf. \cite{Gonzalez-2016}). Furthermore, the limiting behavior at full compaction appears to be $\alpha \rightarrow 2(2+1/m)/3$ for particle-particle interactions, and $\alpha \rightarrow 1+(2+1/m)/3$ for particle-wall interactions. A detailed analysis of the significance of the recurrent term $(2+1/m)/3$ is beyond the scope of this work.

\subsection{Probability distribution of contact radii}

The probability distribution of normalized contact radii $\bar{a}=a/a_{av}$ is determined from the particle mechanics simulations in a manner similar to the one used for building the probability distribution of contact forces. Figure \ref{Fig-ForcesRadii-m=1AND5} shows that the probability distribution evolves during the compaction process and depends on the hardening exponent $m$. The function $P(\bar{a};m)$ should then follow from Eq. \eqref{Eq-Force} using $f \propto a^{2+1/m}$. However, a tight bound for $\bar{a}$ is not available and, following \cite{Gonzalez-2016}, $\bar{f} \le \bar{a}^{2+1/m} \le \bar{a}^3$ is adopted
\begin{equation}
	P(\bar{a};m) 
	\propto 
	\bar{a}^{3(\gamma -1)} 
	\exp
	\left[ 
	-\lambda^\alpha \left|\bar{a}^{3}-a_0\right|^\alpha  
	\right]
\label{Eq-Radius}
\end{equation}
which lacks an explicit dependency on $m$. The parameters $\alpha$, $a_0$, $\gamma$ and $\lambda$ are then determined by fitting Eq. \ref{Eq-Radius} to the particle mechanics simulation, expecting them to depend on $m$. Figure \ref{Fig-ForcesRadii-m=1AND5} shows that the proposed functionality fully-describes each probability distribution with remarkable accuracy. Similar to probability distributions of contact forces, the distributions are similar for particle-particle and particle-wall interactions at jamming onset, for any value of $m$, but become very different at full compaction and depend on the hardening exponent $m$. A simplification similar to \eqref{Eq-p-p-param}-\eqref{Eq-p-w-param} remains an open problem worthy of future studies \cite{Gonzalez-2016}.

\subsection{Granular fabric anisotropy}

\begin{sidewaysfigure}
\centering{
\begin{tabular}{llll}
    \includegraphics[trim= 0mm 0mm 0mm 0mm, clip, scale=0.61]{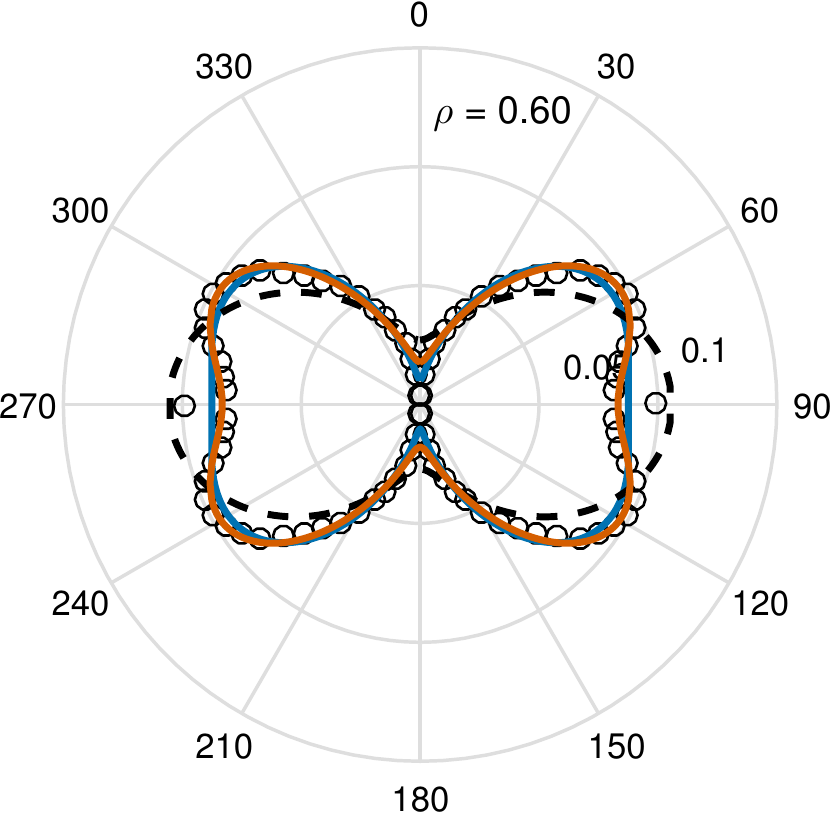} 
	&
    \includegraphics[trim= 0mm 0mm 0mm 0mm, clip, scale=0.61]{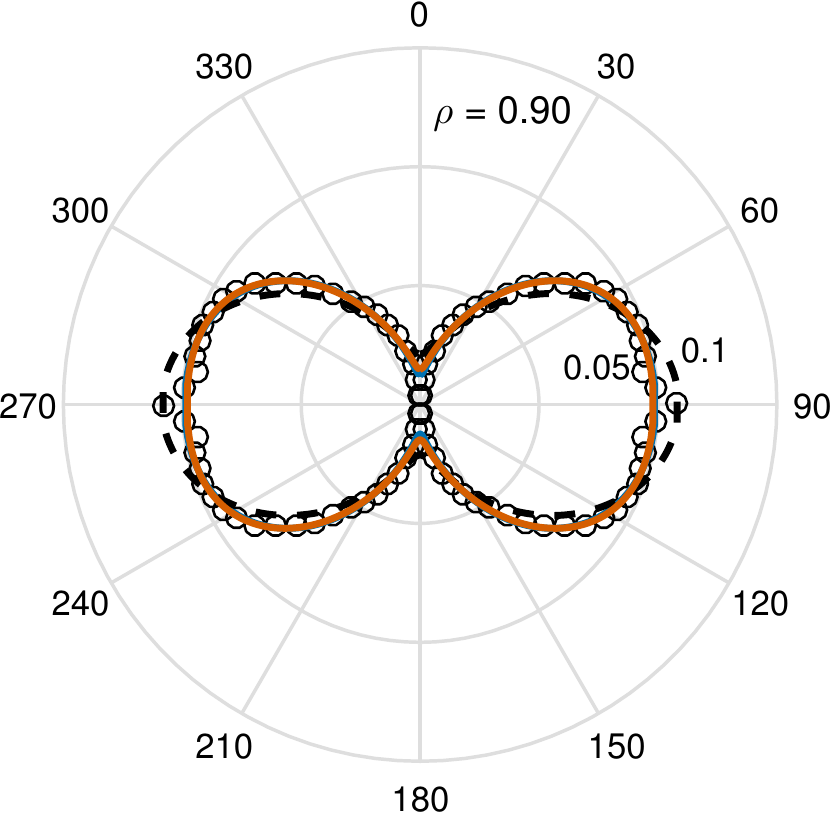}	
	&
    \includegraphics[trim= 0mm 0mm 0mm 0mm, clip, scale=0.61]{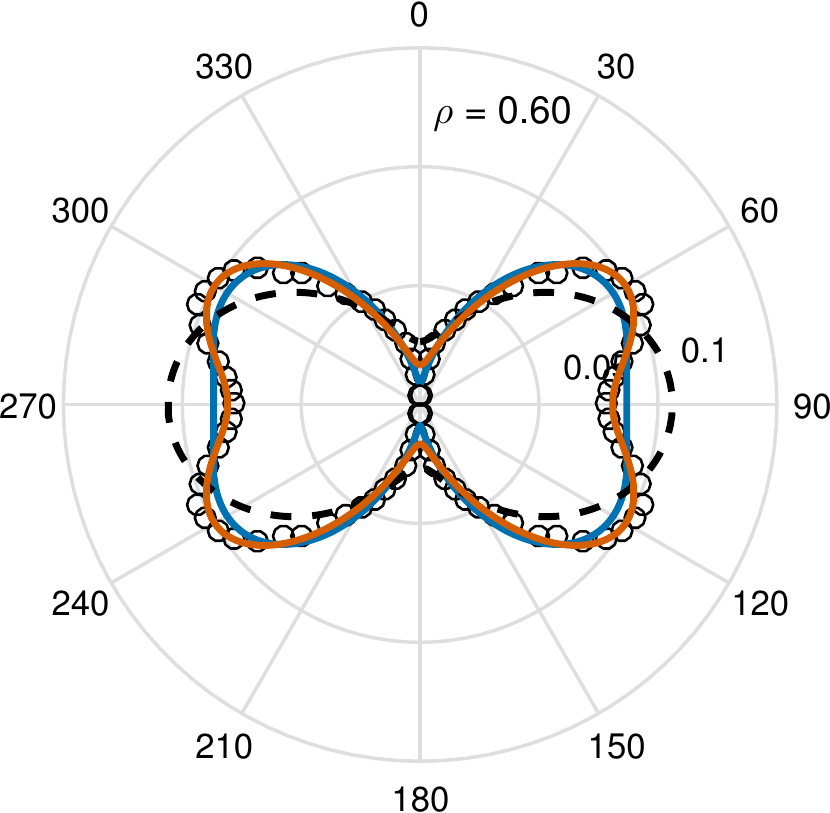} 
	&
    \includegraphics[trim= 0mm 0mm 0mm 0mm, clip, scale=0.61]{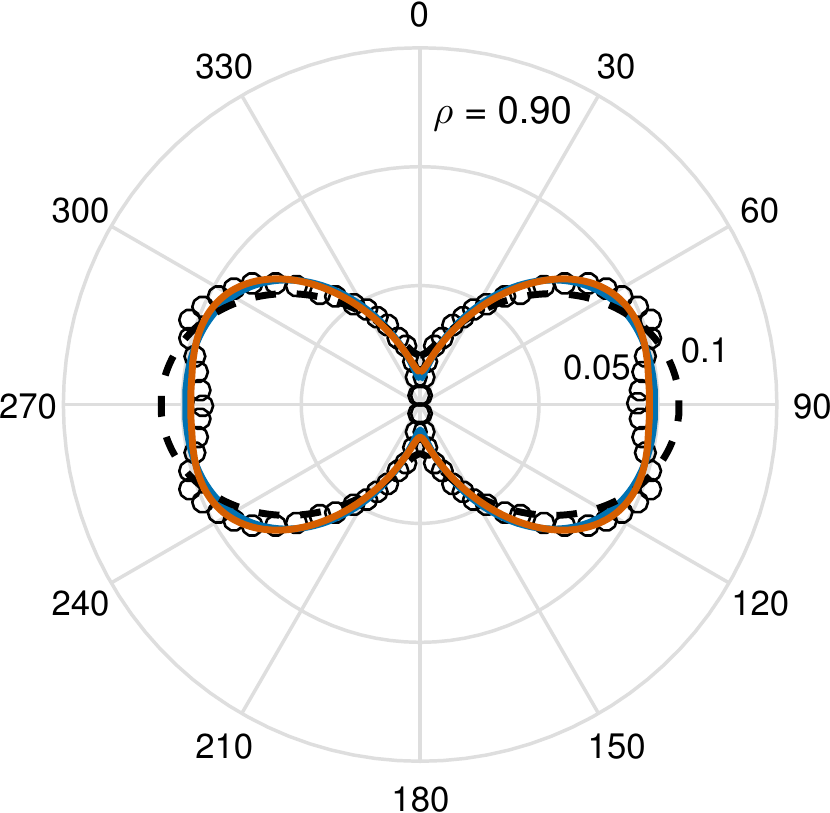}		
	\vspace{-.18in}
\\
	\small{(a)}
	&
	\small{(b)}
	&
	\small{(c)}
	&
	\small{(d)}	
\\
    \includegraphics[trim= 0mm 0mm 0mm 0mm, clip, scale=0.61]{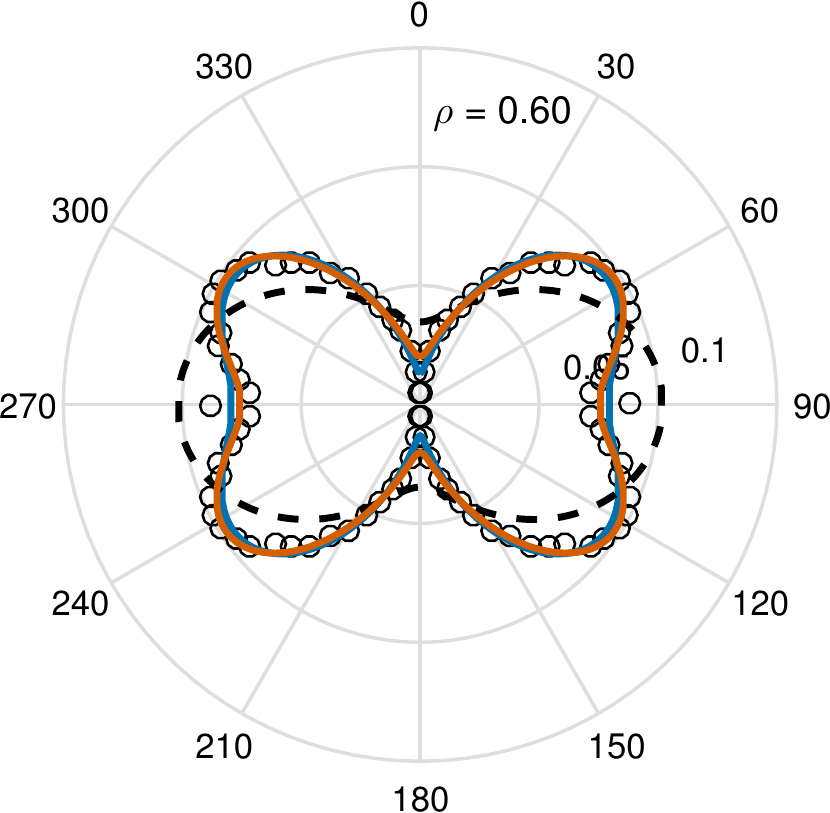} 
    &
    \includegraphics[trim= 0mm 0mm 0mm 0mm, clip, scale=0.61]{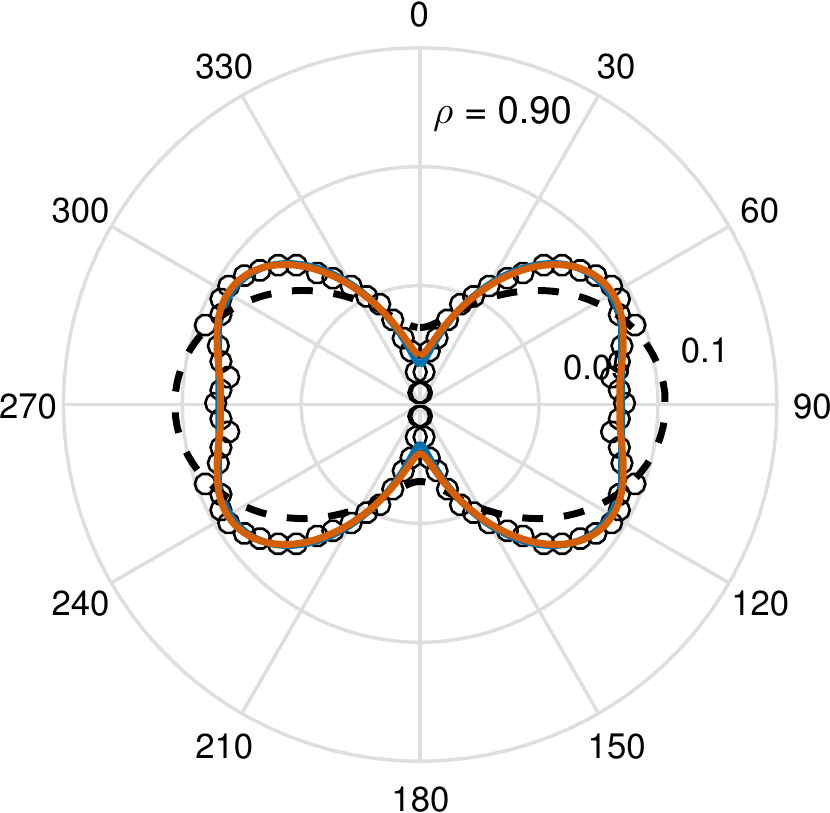}	
	&
    \includegraphics[trim= 0mm 0mm 0mm 0mm, clip, scale=0.61]{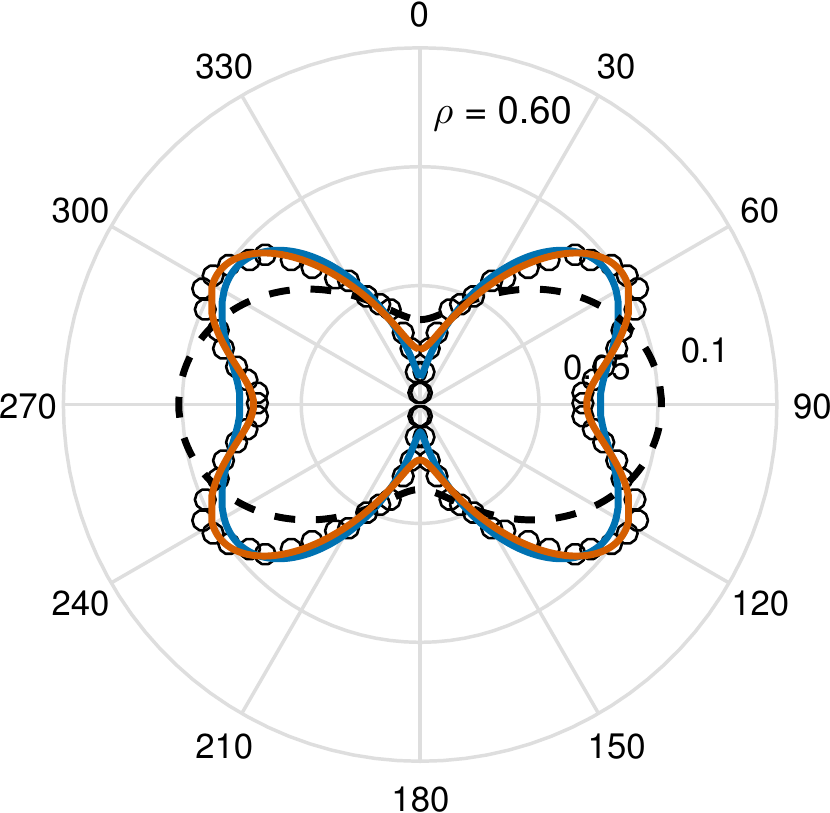} 
    &
    \includegraphics[trim= 0mm 0mm 0mm 0mm, clip, scale=0.61]{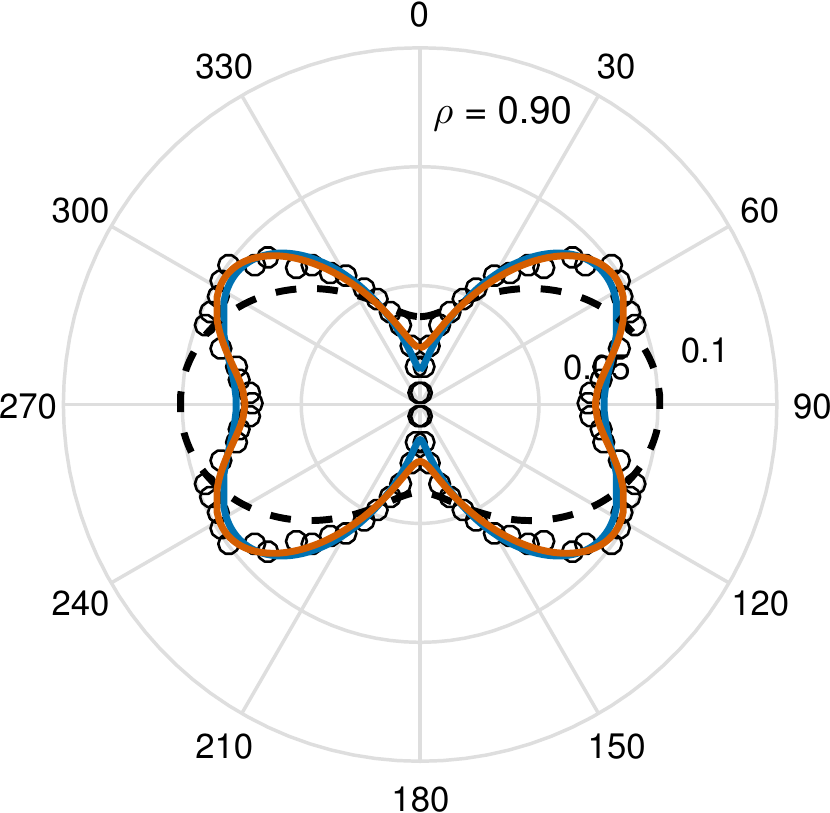}		
	\vspace{-.18in}
\\
	\small{(e)}
	&
	\small{(f)}
	&
	\small{(g)}
	&
	\small{(h)}
\end{tabular}
}
\caption{Directional distribution of contact forces larger than $f_{av}$ ($\circ$, top row) and of all contact forces ($\circ$, bottom row) at two determined from particle mechanics simulations of 40k packings of plastic spheres with hardening exponent $m=1$ (a,b,e,f) and $m=5$ (c,d,g,h). The dashed line corresponds to the best fit of a first order expansion of $\xi$ to the particle mechanics simulation; blue line corresponds to a second order expansion; orange line corresponds to a third order expansion, i.e., to Eq. \eqref{xi-transverse-246}.}
\label{Fig-ForceDist-m=1AND5}
\end{sidewaysfigure}

Material microstructure evolves in an anisotropic manner during loading \cite{Gonzalez-2016,TordesillasForceChain}. Specifically, during die-compaction, the axis of cylindrical die acts as an axis of rotational symmetry transforming the granular assembly into a transversely isotropic material \cite{Malvern-Continuum}. Statistics of the microstructural features are therefore, independent of the azimuth angle $\phi$, and are functions only of the polar angle measured from the axis of rotational symmetry $\theta$. Hence, the directional density distribution of microstructural features can be expressed by a spherical harmonic expansion using Legendre polynomials in terms of $\cos\theta$ \cite{Payam-Gibbs,Kanatani,Chang-Packing}. The present study indicates that the first three terms in the distribution function suffice for representing the directional distribution of contact forces in the packing, that is
\begin{align}
	\xi\left(\theta,\phi\right)
	= & 
	\dfrac{1}{4\pi} 
	\Bigg\{
		1+\frac{1}{2}a_{20}\left[3\cos^2\theta+1\right] 
		\label{xi-transverse-246} \\
	&
		+\frac{1}{8}a_{40}\left[35\cos^4\theta-30\cos^2\theta+3\right] \nonumber \\
	& 	
	 	+\frac{1}{16}a_{60}\left[231\cos^6\theta-315\cos^4\theta+105\cos^2\theta-5\right]
	\Bigg\} \nonumber
\end{align}
where $a_{k0}$ represent fabric parameters. Figure \ref{Fig-ForceDist-m=1AND5} shows that the above equation accurately describes the directional distribution of particle-particle contact forces, at two levels of compaction (namely at the onset of jamming and near full compaction) and for plastic spheres with two different hardening exponents $m$ (namely $m=1$ and $m=5$). It is evident from the figures that a first order expansion of $\xi$ does not adequately describe the particle mechanics simulation results. 

\begin{figure*}[htb]
\centering{
\begin{tabular}{ll}	
	\includegraphics[trim= 10mm 0mm 54mm 0mm, clip, scale=0.62]{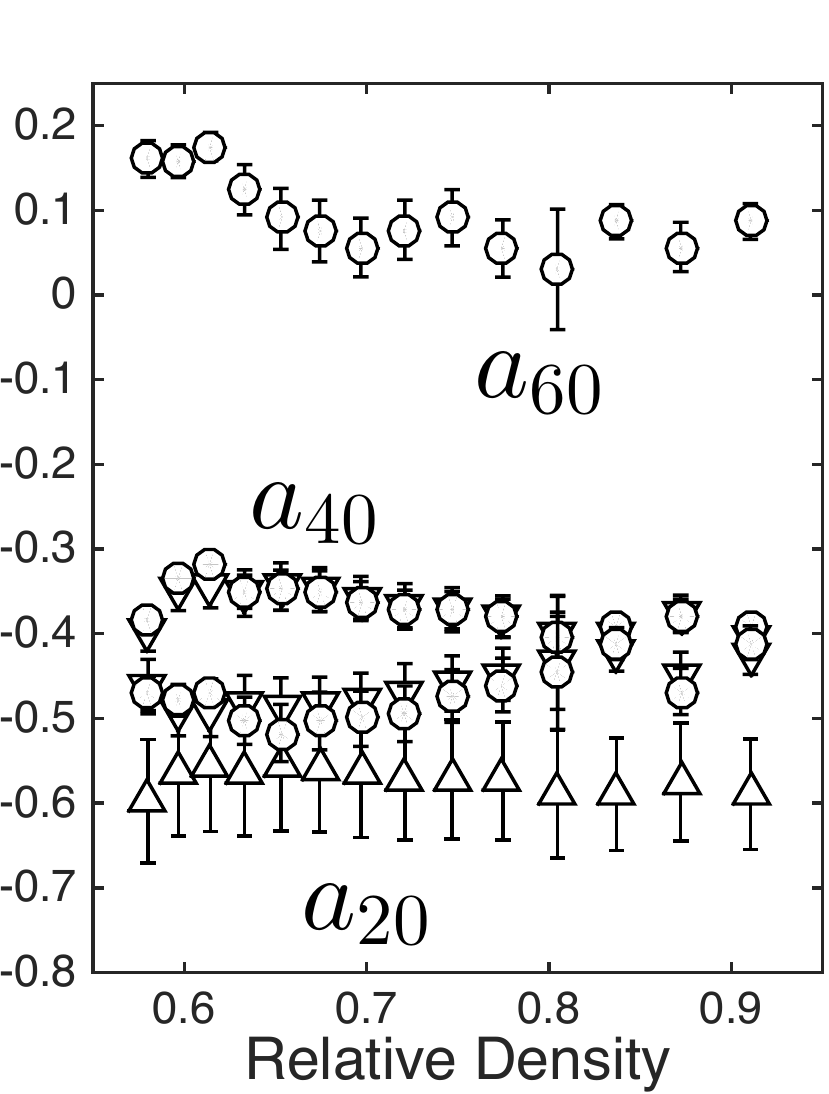}
	&
	\includegraphics[trim= 10mm 0mm 54mm 0mm, clip, scale=0.62]{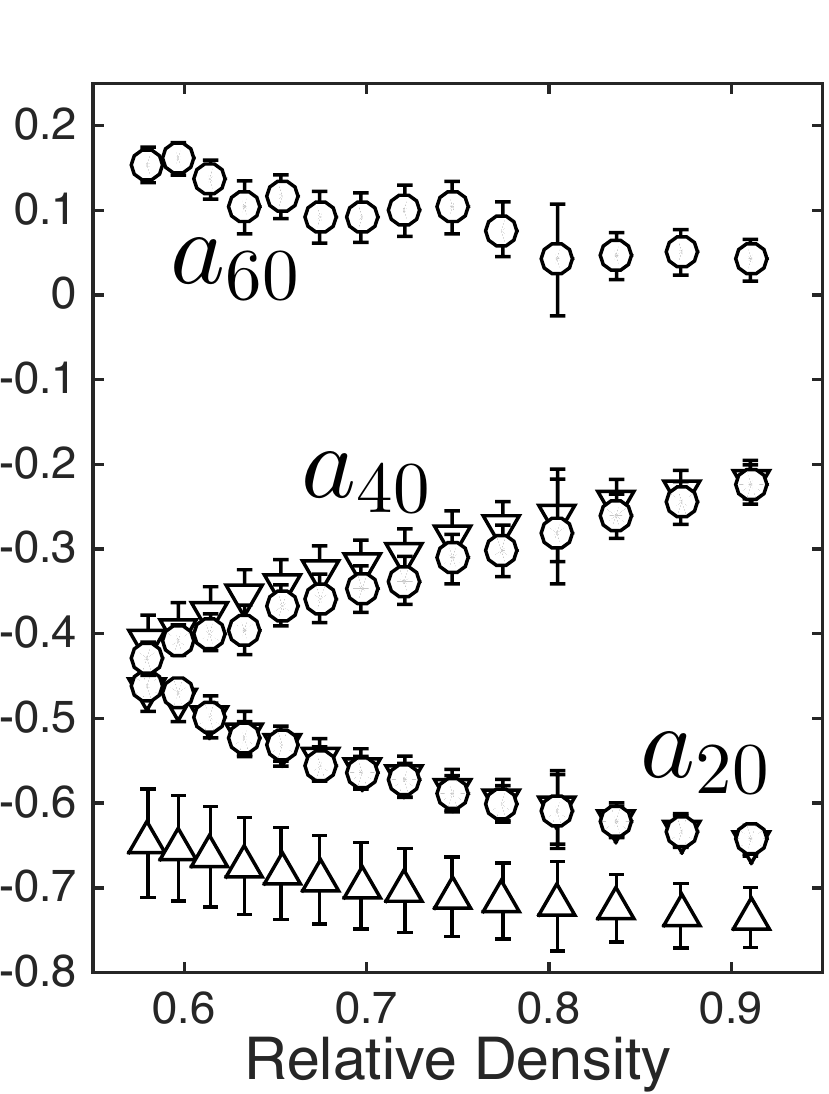}
	\vspace{-.18in}
\\
	\small{(a)}
	&
	\small{(b)}
\end{tabular}
\vspace{-.1in}
\begin{tabular}{ll}	
	\includegraphics[trim= 10mm 0mm 54mm 0mm, clip, scale=0.62]{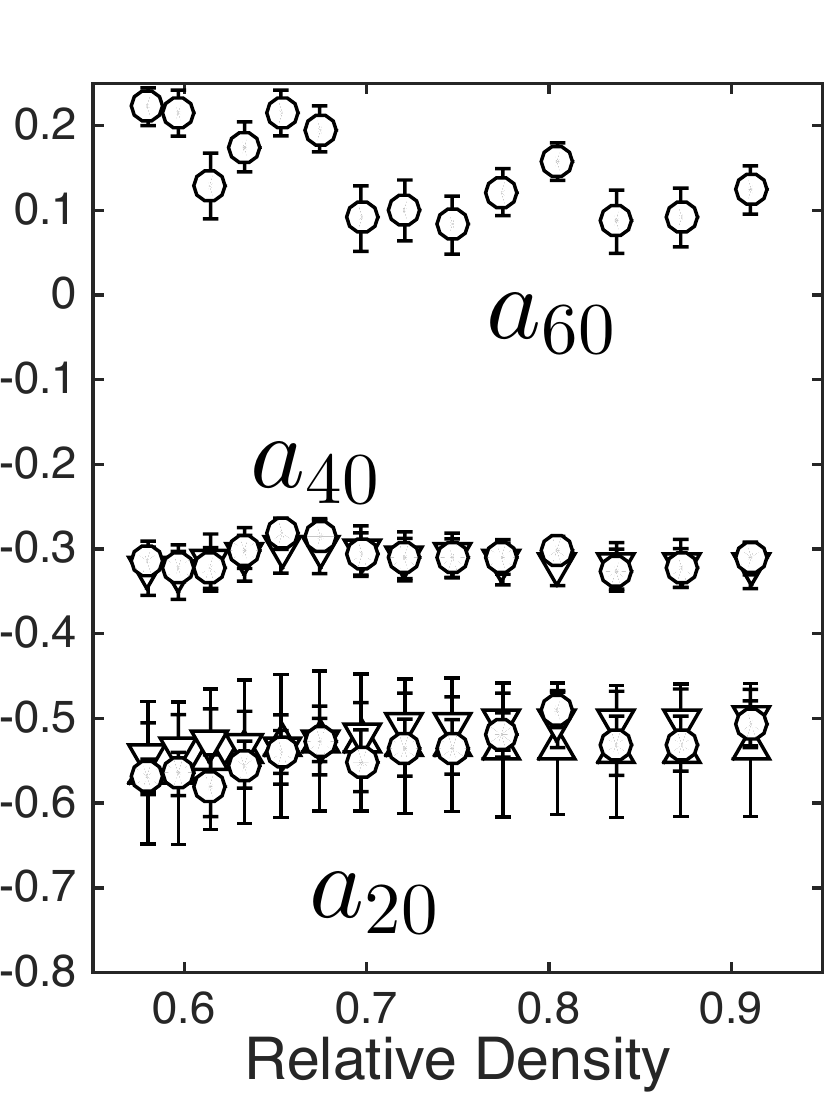}
	&
	\includegraphics[trim= 10mm 0mm 54mm 0mm, clip, scale=0.62]{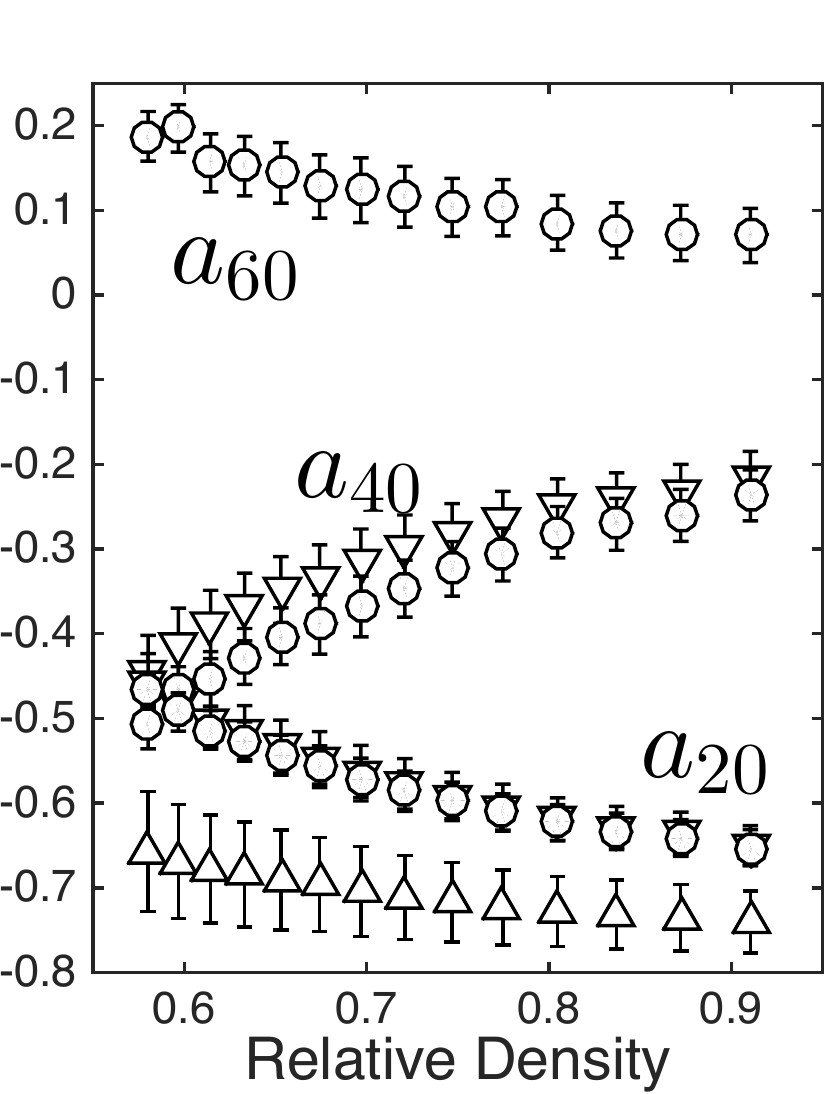}	
	\vspace{-.18in}
\\
	\small{(c)}
	&
	\small{(d)}		
\end{tabular}
}
\caption{Evolution of fabric parameters governing the directional distribution of strong contact forces (a,c) and all contact forces (b,d) for packings of plastic particles with hardening exponent $m=1$ (a,b) and $m=5$ (c,d). First ($\vartriangle$), second ($\triangledown$) and ($\circ$) order approximations of $\xi$.}
\label{Fig-a204060-m=1AND5}
\end{figure*}

For completeness, Figure \ref{Fig-a204060-m=1AND5} shows the evolution of $a_{20}$, $a_{40}$ and $a_{60}$, for a first ($a_{20}\ne 0$), second ($a_{20}\ne 0$, $a_{40}\ne 0$) and third ($a_{20}\ne 0$, $a_{40}\ne 0$, $a_{60}\ne 0$) order expansions of the distribution function for both networks of all contact forces and strong contact forces, that is of $f > f_{av}$. A clear evolution of fabric parameters during the compaction process is evident. Interestingly, these results suggest that the algebraic tail of $P(\bar{f};m)$ (Figure \ref{Fig-p-p-parameters}) and its fabric parameters (Figures \ref{Fig-a204060-m=1AND5}(a)(c)) depend on $m$, while the fabric parameters of all $\bar{f}$ do not depend strongly on $m$ (Figures \ref{Fig-a204060-m=1AND5}(b)(d)). A systematic investigation of this observation is a worthwhile direction of future research. 

\section{Concluding Remarks}

An extensive campaign of numerical experiments, simulating the die compaction of monodisperse plastic spheres that exhibit power-law plastic hardening behavior were performed \cite{nanoHUBTool}. The results of these simulations were analyzed to map the relationship between particle plastic properties, loading conditions, and statistical features of the resulting microstructure. The findings of the study are summarized below.
\begin{enumerate}
\itemsep0em
\item The jamming point does not vary significantly with material properties and it is rather a geometric property of the packing, with a weak dependency on the scale factor $D/R$.
\item  The critical exponent that describes the evolution of mean coordination number during compaction does not exhibit dependency on material properties, while the coordination number at the onset of jamming exhibits a weak dependency on hardening exponent.
\item The exponent and stiffness of the power law that describes the evolution of  applied compaction pressure depend nonlinearly on hardening exponent, and only weakly on particle size and scale factor $D/R$. These values exhibit an asymptote at $m \rightarrow \infty$, and are consistent with those of elastic particles at $m=1$. The stiffness of the compaction pressure power law depends linearly on material power law stiffness.
\item The probability distribution of normalized particle-particle and particle-wall contact forces are similar at the onset of jamming, but evolve differently and depend on hardening exponent. The tails of both distributions transition from exponential ($\alpha=1$) to algebraic; with limiting behavior at full compaction equal to $\alpha \rightarrow 2(2+1/m)/3$ for particle-particle interactions, and equal to $\alpha \rightarrow 1+(2+1/m)/3$ for particle-wall interactions.
\item The directional distribution of contact forces can be described by the first three terms of a harmonic expansion. The fabric parameters of the network of strong forces depend on the hardening exponent, while those of all contact forces do not.
\end{enumerate}

\section*{Acknowledgments}
The motivation for this work arose from a special topics course for undergraduate students, ME479, on microstructure evolution of granular systems given as an {\it independent research project} from the spring of 2014 to the spring of 2017 at the School of Mechanical Engineering at Purdue University. Participants were Kiran Balakrishnan (Spring 2014), Jili Liu (Summer 2014 to Fall 2014) and Alex Thomas (Spring 2016 to Spring 2017). The authors gratefully acknowledge the support received from the National Science Foundation grant number CMMI-1538861, from Purdue University's startup funds, and from the Network for Computational Nanotechnology (NCN).

\bibliographystyle{IEEEtran}
\bibliography{IEEEabrv,MIMO}

\end{document}